\documentclass[conference, 9pt]{IEEEtran}
\IEEEoverridecommandlockouts

\usepackage{cite}
\usepackage{amsmath,amssymb,amsfonts}
\usepackage{algorithmic}
\usepackage{graphicx}
\usepackage{textcomp}
\usepackage{xcolor}
\usepackage{hyperref}
\usepackage{comment}
\usepackage{multirow, multicol} 
\usepackage{booktabs, balance}
\usepackage{makecell}
\usepackage[colorinlistoftodos]{todonotes}
\def\BibTeX{{\rm B\kern-.05em{\sc i\kern-.025em b}\kern-.08em
    T\kern-.1667em\lower.7ex\hbox{E}\kern-.125emX}}

\begin{document}


\title{WeDefense: \\
A Toolkit to Defend Against Fake Audio
}
\IEEEspecialpapernotice{Preprint notice: This preprint corresponds to the version completed by June 4, 2025 and previously submitted to ASRU 2025. A revised version is in progress.}

\author{
\IEEEauthorblockN{
Lin Zhang\IEEEauthorrefmark{1},
Johan Rohdin\IEEEauthorrefmark{2},
Xin Wang\IEEEauthorrefmark{3},
Junyi Peng\IEEEauthorrefmark{2},
Tianchi Liu\IEEEauthorrefmark{4},
You Zhang\IEEEauthorrefmark{5},\\
Hieu-Thi Luong\IEEEauthorrefmark{6},
Shuai Wang\IEEEauthorrefmark{7},
Chengdong Liang\IEEEauthorrefmark{8},
Anna Silnova\IEEEauthorrefmark{2},
Nicholas Evans\IEEEauthorrefmark{9}
}
\IEEEauthorblockA{\IEEEauthorrefmark{1} Johns Hopkins University, \IEEEauthorrefmark{2} Brno University of Technology, \IEEEauthorrefmark{3} National Institute of Informatics}
\IEEEauthorblockA{\IEEEauthorrefmark{4} National University of Singapore, \IEEEauthorrefmark{5} University of Rochester, \IEEEauthorrefmark{6} Nanyang Technological University}
\IEEEauthorblockA{\IEEEauthorrefmark{7} Nanjing University, \IEEEauthorrefmark{8} WeNet Opensource Community, \IEEEauthorrefmark{9} EURECOM}
}

\maketitle

\begin{abstract}
The advances in generative AI have enabled the creation of synthetic audio which is
perceptually indistinguishable from real, genuine audio. Although this stellar progress enables many positive applications, it also raises risks of misuse, such as for impersonation, disinformation and fraud.
Despite a growing number of open-source fake audio detection codes released through numerous challenges and initiatives, most are tailored to specific competitions, datasets or models. 
A standardized and unified toolkit that supports the fair benchmarking and comparison of competing solutions with not just common databases, protocols, metrics, but also a shared codebase, is missing.
To address this, we propose WeDefense, the first open-source toolkit to support both fake audio detection and localization.
Beyond model training, WeDefense emphasizes critical yet often overlooked components: flexible input and augmentation, calibration, score fusion, standardized evaluation metrics, and analysis tools for deeper understanding and interpretation.
The toolkit is publicly available with interactive demos for fake audio detection and localization\footnote{\url{https://huggingface.co/spaces/wedefense/fake_audio_localization_demo}}.

\end{abstract}


\section{Introduction}

The astonishing progress in generative AI has led to the development of systems that make it increasingly straightforward to synthesize audio. This is ever more challenging to distinguish from real, genuine audio. Although a plethora of generative audio project (such as Amphion\footnote{\url{https://github.com/open-mmlab/Amphion}}~\cite{zhang2024amphion}, CosyVoice\footnote{\url{https://github.com/FunAudioLLM/CosyVoice}}~\cite{du2025cosyvoice3}, WhisperSpeech\footnote{\url{https://github.com/WhisperSpeech/WhisperSpeech}}, PlayDiffusion\footnote{\url{https://github.com/playht/PlayDiffusion}}, VoiceSculptor), have emerged in recent years, efforts to develop well-supported, standardized toolkits for the detection and localization of fake audio have lagged behind.
To promote broader research into the development of defending against fake audio, we introduce the open-source, community-led \textbf{WeDefense}\footnote{\url{https://github.com/zlin0/wedefense}} project and toolkit. 

Existing open-source projects for fake audio detection and fake audio localization typically suffer from limited flexibility and poor scalability.
For example, various individual implementations proposed for fake audio detection such as 
RawNet2\footnote{\url{https://github.com/eurecom-asp/rawnet2-antispoofing}}~\cite{tak2021rawnet2}, AASIST\footnote{\url{https://github.com/clovaai/aasist}}~\cite{jung2022aasist}, XLSR-SLS\footnote{\url{https://github.com/QiShanZhang/SLSforASVspoof-2021-DF}}~\cite{zhang2024sls}, and Nes2Net\footnote{\url{https://github.com/Liu-Tianchi/Nes2Net}}~\cite{liu2025nes2net}, all released alongside research papers, are usually maintained as standalone projects, each one with a focus on different data processing techniques and specific models. 
The baselines provided through the ASVspoof challenge\footnote{\url{https://github.com/asvspoof-challenge}} series~\cite{asvspoof2021, wang24asvspoof5} are typically task- and dataset-specific, lacking generalizability and flexibility for broader application. Moreover, since the baselines are developed by different organizers, there is considerable variability in the code structure. 

Although codebases like project-NN-PyTorch\footnote{\url{https://github.com/nii-yamagishilab/project-NN-Pytorch-scripts}}~\cite{wang2021comparative} consolidate several deepfake detection models, they were designed originally for other tasks and lack a broad coverage of models proposed by different teams.
The recently proposed SONAR\footnote{\url{https://github.com/Jessegator/SONAR}} benchmark~\cite{li2024sonar} integrates several models into the AASIST codebase. However, its foundation upon the original AASIST structure limits the flexibility for the incorporation of alternative model architectures. 

 
All the aforementioned codebases are tailored primarily for fake audio detection. However, for fake audio localization that locates the fake regions within an audio, only a limited number of codebases are publicly available. These are typically released alongside specific models such as the SSL-gMLP-based multi-resolution model\footnote{\url{https://github.com/nii-yamagishilab/PartialSpoof}}~\cite{zhang2022partialspoof}, TDL\footnote{\url{https://github.com/xieyuankun/TDL-ADD}}~\cite{xie2024TDL}, BAM\footnote{\url{https://github.com/media-sec-lab/BAM}}~\cite{zhong2024BAM}, and CFPRF\footnote{\url{https://github.com/ItzJuny/CFPRF}}~\cite{wu2024CFPRF}.

Though toolkits such as Kaldi\footnote{\url{https://github.com/kaldi-asr/kaldi}}~\cite{povey2011kaldi}, icefall\footnote{\url{https://github.com/k2-fsa/icefall}}, ESPnet\footnote{\url{https://github.com/espnet/espnet}}~\cite{watanabe18espnet}, and SpeechBrain\footnote{\url{https://speechbrain.github.io/}}~\cite{ravanelli2021speechbrain} have been adopted widely in the speech processing community, they were initially designed for speech recognition. Although they were all later extended to other tasks, none covers fake audio detection or localization. 
Furthermore, they are often heavy-weight, with extensive functionalities for broader speech applications, many of which are unnecessary for defensing against fake audio.
The existing toolkits, most characterized by other task-specific design choices, lack support for variable-length inputs, data augmentation, and acoustic features designed specifically for fake audio detection. Finally, none can be readily adapted to tackle the localization of short fake audio segments embedded within otherwise bona fide recordings.
%
Without a common toolkit designed specifically for fake audio detection and localization\footnote{As of January 2026, we became aware of \href{https://github.com/Yaselley/deepfense-framework}{DeepFense}, another publicly available toolkit primarily focused on deepfake detection.}, it remains somewhat difficult to compare solutions on an entirely level playing field. The implementation of different solutions within a common codebase, in addition to the use of common databases, protocols and metrics, is the next step toward fairer benchmarking and analyses. 

WeDefense is a relatively light-weight, modular, and extensible toolkit tailored to defend against fake audio. The current version targets detection and localization, while additional functionality is planned and already under development. We hope that the WeDefense project will attract additional contributors and users and evolve into a widely adopted resource, one that supports reproducible research and keeps pace with advances in generative speech technologies.
\begin{figure*}[!tb]
\centerline{\includegraphics[width=0.99\linewidth]{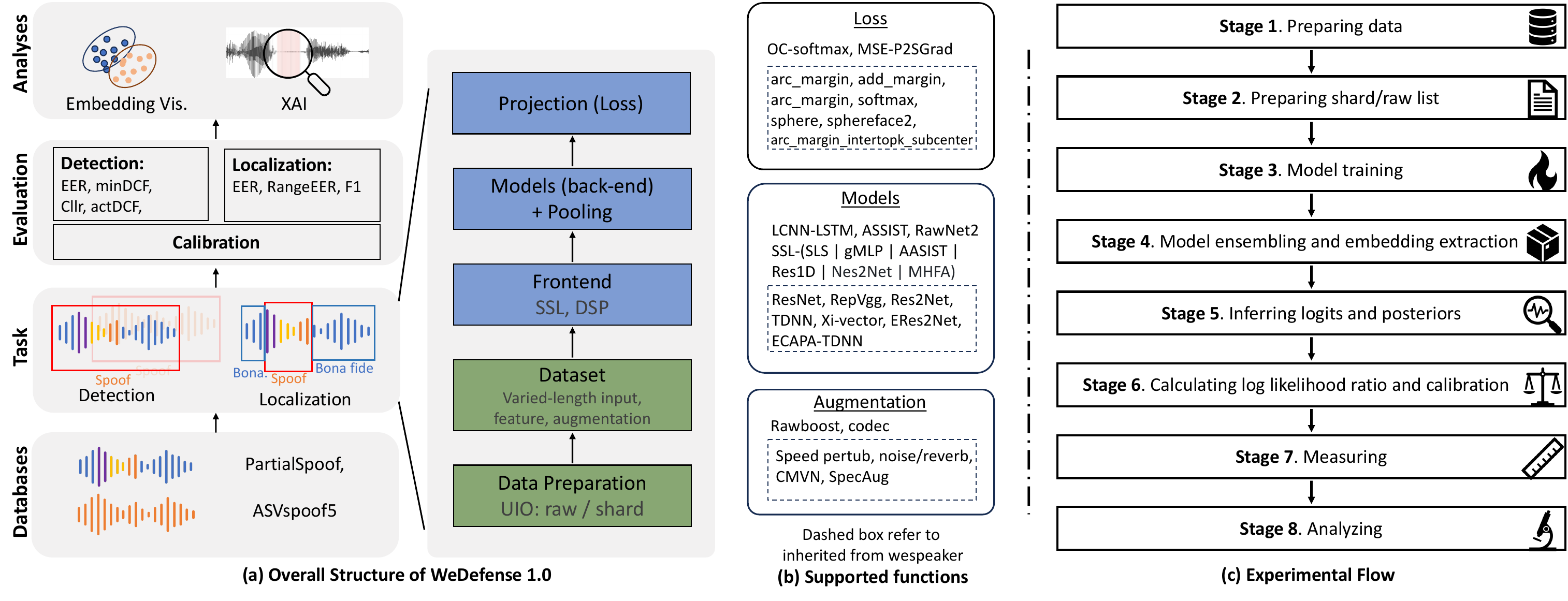}}
\caption{Overall Structure of WeDefense.}\label{fig:structure}
\end{figure*}
%
%
The WeDefense project follows WeNet\footnote{\url{https://github.com/wenet-e2e/wenet}}~\cite{zhang2022wenet} and WeSpeaker\footnote{\url{https://github.com/wenet-e2e/wespeaker}}~\cite{wang2023wespeaker}.
The key features of the WeDefense toolkit are described in the following.
\begin{itemize}
    \item \textbf{First open-source toolkit for fake audio detection and localization.} WeDefense is not restricted to any specific dataset, and is easily adaptable to other general 
    datasets. It contains implementations of mainstream models and is designed to be extensible to incorporate more powerful models in the future.
    \item \textbf{Unified design across models and datasets.} WeDefense integrates and unifies configurations and setups from existing open-source codes, which were often developed for a specific competition/dataset or model, into a more general and maintainable structure.
    
    \item \textbf{Designed specifically for fake audio detection.} The toolkit includes features which are specific to the study of fake audio detection, such as the treatment of variable-length inputs and common data augmentations (e.g., RawBoost\footnote{\url{https://github.com/TakHemlata/RawBoost-antispoofing}}~\cite{Tak2021RawBoost:Anti-Spoofing} and codec simulation~\cite{rohan2021augmentation_codec}). It also includes widely used techniques and tricks borrowed from the automatic speaker verification field, such as model averaging for stable and reliable performance estimation.
    \item \textbf{Support for analyses and evaluation with postprocessing.} WeDefense integrates several processing and analysis tools to support calibration, 
    embedding visualization, and explainable AI, to enable deeper exploration, evaluation and understanding of model behavior.
    \item \textbf{Modular and extensible design.} WeDefense retains modular interfaces, making the toolkit easily extendable to other tasks related to fake audio detection in the future.
    \item \textbf{Deployment-ready.} 
    WeDefense models can be exported via TorchScript (JIT) for deployment. Pretrained models and example C++ deployment code are also provided to support integration and real-world applications. 
\end{itemize}

\section{Fake Audio Detection and Localization}
WeDefense supports two common tasks in defending against fake audio, as shown in the ``Task'' part of Figure~\ref{fig:structure} (a). 


\subsection{Fake Audio Detection}
Fake audio detection aims to detect whether an entire audio sample is real or fake. This task has attracted researchers' attention for over a decade, particularly since the ASVspoof challenge series began~\cite{evans2013spoofing, Wu2014}.
Recently, the task has become even more challenging with the emergence of a new scenario called ``Partial Fake''\footnote{The original term used in the literature is ``Partial Spoof''; we adopt ``Partial Fake'' here to reflect a broader range of manipulations beyond spoofing specifically for authentic systems.}~\cite{Zhang2021PartialSpoof, Yi2021halftruth}, where only a portion of the audio is generated while the rest remains real. Such cases are even more difficult to detect for both humans and machines\cite{alali2025partial}, as the real part could mislead the detector. The core objective of detecting fake audio remains the same in both full and partial fake scenarios, but many methods developed for fully spoofed samples overlook their performance on the partial fake case, likely for the same reasons mentioned earlier in the introduction. WeDefense, as a unified platform, is built around task-oriented rather than data-specific design principles. This project aims to encourage researchers to develop, evaluate, and compare methods across a wide range of databases in a consistent and extensible framework.

\subsection{Fake Audio Localization}
Fake audio localization aims to locate the fake regions within an audio that has been manipulated. This task has emerged as a critical response to the increasingly common ``Partial Fake'' scenario. Accurately locating these fake regions enables deeper analysis of the spoofed content and may help uncover the attacker’s intent. Furthermore, isolating the real (unaltered) regions could potentially allow for reconstruction of the original utterance~\cite{zhang2024partialspoof_thesis}. 

\section{WeDefense}\label{sec:wedefense}
\subsection{Overall Structure}

Figure~\ref{fig:structure}(a) illustrates the architecture and core functionalities supported by the current version of WeDefense. It supports two primary tasks: fake audio detection and localization. In addition, it integrates tools for calibration and widely used evaluation metrics. The toolkit also integrates analysis tools, including embedding visualization (UMAP)~\cite{mcinnes2018umap-software}, explainable AI (XAI) by Grad-CAM~\cite{selvaraju2017grad}.

Figure~\ref{fig:structure}(b) presents the supported loss functions, models, and augmentation techniques in WeDefense. 

Figure~\ref{fig:structure}(c) shows the experimental workflow of a standard WeDefense recipe. When a new dataset is introduced, users can easily adapt existing examples to conduct complete experiments by following the structured stages shown.

We will introduce more details in the following subsections.


\subsection{Databases - focus on both partially and fully spoofed audio}

All models and functions in WeDefense can be provided to any database if dataset-specific configuration files can be prepared. We currently provided examples for two databases.

\begin{itemize}
    \item \textbf{PartialSpoof}~\cite{Zhang2021PartialSpoof}: The first publicly available dataset designed for the Partial Fake scenario. It has been downloaded more than 7,000 times by June 2025. PartialSpoof is constructed based on ASVspoof2019~\cite{Wang2020data} by randomly substituting speech segments between different classes while maintaining the same speaker identity. It includes fake segments generated by 17 different TTS/VC systems.
    
    \item \textbf{ASVspoof5}~\cite{wang2025ASVspoof5_database}: The latest database used in the ASVspoof challenge series. It is built from crowdsourced data, and comprises over 2,000 speakers and 32 algorithms considering adversarial attacks.
\end{itemize}

\subsection{Input and Augmentation - flexible for defending against fake audio}
A commonly used approach for training classification models is to use fixed-length inputs, often by trimming or padding the audio. Although this is simple and widely adopted in fake audio detection, it may overlook information, especially in partially fake audio, in which the short fake segment might be trimmed off, but the label is still ``fake''. Recent studies have shown that using varied-length inputs can be beneficial in certain scenarios~\cite{wang2021comparative}, making this strategy particularly important to consider. Thus, WeDefense integrates support for varied-length input.

Moreover, although recent work increasingly focuses on self-supervised learning (SSL) model~\cite{wang2021ssl}, audio large language model~\cite{gu2025LLM4ADD}, and other raw waveform-based end-to-end models, traditional acoustic features remain useful, especially for lightweight models. Therefore, WeDefense also supports widely used acoustic features such as filter bank (Fbank) and linear frequency cepstrum coefficients (LFCC)~\cite{davis_comparison_1980}.


As fake audio generation technologies continue to evolve, it is a common strategy to improve model generalization by expanding the domain coverage of training data. Therefore, data augmentation techniques are widely adopted. Currently, WeDefense supports several commonly used augmentations in speech and audio processing, including SpecAugment~\cite{park19specaug}, MUSAN~\cite{snyder2015musan}, and RIR-based augmentation, and speed perturbation~\cite{povey2011kaldi}. In addition, it also incorporates augmentation methods specifically proposed for fake audio detection, such as RawBoost~\cite{Tak2021RawBoost:Anti-Spoofing} and codec-based augmentation~\cite{rohan2021augmentation_codec}.

\subsection{Models - support both conventional models and SSL-based models, detection and localization tasks.}

For fake audio detection, WeDefense includes several models for deepfake audio detection. At the time of submitting this paper, it includes widely used convolutional neural network (CNN), Light CNN-LSTM~\cite{wang2021comparative}, ResNet, and several recently popular SSL-based models: those two used in the ASVspoof challenge: SSL-AASIST~\cite{tak22_odysseyw2v2}, SSL-MHFA~\cite{Peng2023AnVerification}.
Two introduced for PartialSpoof database: SSL-gMLP~\cite{zhang2022partialspoof}, SSL-Res1D~\cite{liu24grad_cam}
Top-performend\footnote{\url{https://huggingface.co/spaces/Speech-Arena-2025/Speech-DF-Arena}} open-sourced model: SSL-SLS~\cite{zhang2024sls}. SSL models are loaded through s3prl~\cite{yang21c_interspeech_superb}.
In addition, WeDefense is fully compatible with WeSpeaker, allowing researchers to quickly apply a wide range of models those widely used in the speaker verification task as shown in dashed box in Figure~\ref{fig:structure}(b)

WeDefense also supports fake audio localization, and mainly focuses on uniform segmentation that is trained on frame-level prediction, whose model can be implemented by removing the pooling layer from the detection model. 
Besides those binary frame-level prediction models, WeDefense also integrates BAM, which incorporates a module for boundary detection~\cite{zhong2024BAM}.

\subsection{Loss Functions - including losses specific for defending against fake audio}
As a toolkit for defending against fake audio, WeDefense integrates some losses introduced for fake audio detection: OC-softmax~\cite{zhang2020one} and MSE-P2SGrad~\cite{wang2021comparative}. In addition, WeDefense supports the standard softmax cross-entropy loss and losses proposed for speaker identification but usable for fake audio detection, such as A-softmax~\cite{liu2017sphereface, huang2018angular}, AM-softmax~\cite{wang2018additive}, and AAM-softmax~\cite{deng2019arcface}.

\subsection{Calibration and Fusion} 

Calibration has been widely studied in speaker verification tasks~\cite{brummer2010measuring}, but it has received limited attention in the context of fake audio detection. Poor calibration means that the information provided by the scores is misleading (loosely speaking, the scores are systematically too high or too low), which may lead to suboptimal accept/reject decisions~\cite{brummer2006application}.
This aspect was recently emphasized by the ASVspoof 2024 challenge~\cite{wang24asvspoof5}, which encouraged the use of calibrated scores. To support this, WeDefense integrates the score calibration functionality and currently supports the standard affine transformation calibration learned by logistic regression (LR)~\cite{4291590}.

Fusion is a commonly used strategy in fake audio detection (as well as many other machine learning applications), in which scores from an ensemble of different models are combined into a single score. WeDefense currently integrates two commonly used score-level fusion methods: score average fusion (after normalizing the scores to be in the $[0,1]$ range) and LR-based fusion \cite{4291590}. The LR-based fusion is the same as the LR-based calibration except that the learned affine transformation takes more than one score as input. As a consequence, scores fused in this way are calibrated.

\subsection{Metrics}
\subsubsection{Detection}
To evaluate fake audio detection performance, Equal Error Rate (EER) has long been favored due to its threshold-free nature. More recently, ASVspoof5 has adopted the minimum Detection Cost Function (minDCF) as the primary metric, and further encourages the use of the Cost of Log-Likelihood Ratio ($C_\text{llr}$) and actual DCF to assess calibrated systems. WeDefense integrates all of these evaluation metrics.

\begin{figure*}[!h]
    \centering
    \centerline{\includegraphics[width=0.8\linewidth]{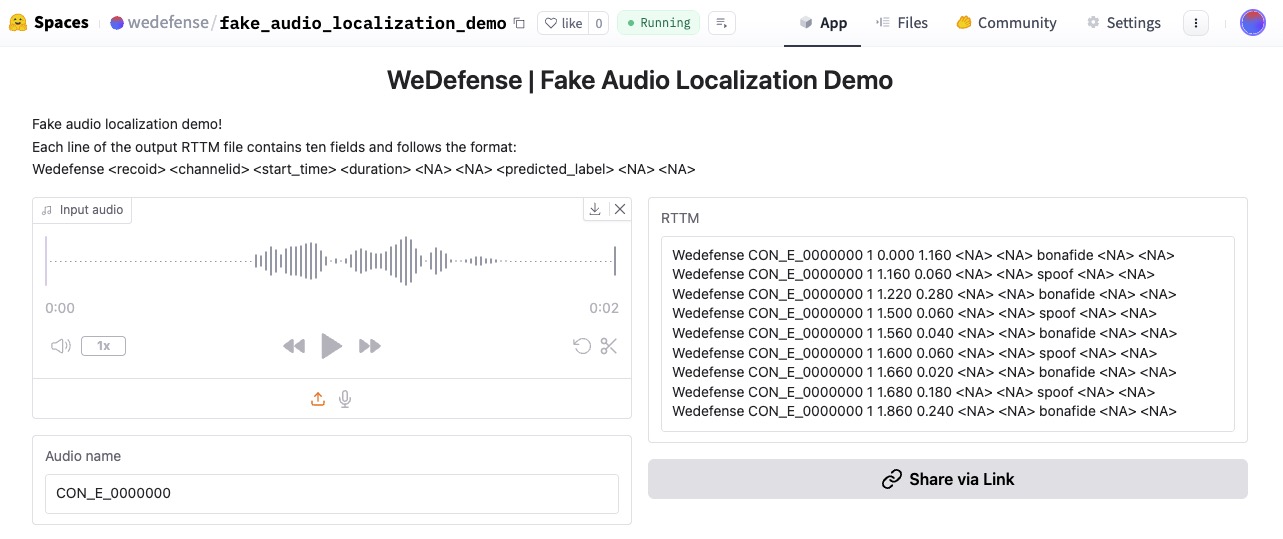}}
    \caption{Interaction Demo 
 of Fake Audio Localization in WeDefense.}
    \label{fig:demo_localization}
\end{figure*}

\subsubsection{Localization}
To measure performance on locating fake segments within partially fake audio, we integrate point-based EER into WeDefense. 
This metric provides a discrete evaluation by comparing frame-level predictions and is efficient to compute. Additional localization metrics are currently under development for integration.

\subsection{Analyses}
A small error rate does not necessarily indicate good performance. Due to the security-sensitive nature of fake audio detection, WeDefense emphasizes thorough analysis besides quantitative metrics.

Embedding visualization helps researchers explore the feature space and better understand model behavior, and has been widely used in fake audio detection studies~\cite{rohdin24_asvspoof}. WeDefense supports embedding visualization using UMAP~\cite{mcinnes2018umap-software}.

In addition, explainable AI is important to understand the mechanism of the fake audio detectors. For example, Liu et al.~\cite{liu24grad_cam} utilized Grad-CAM to interpret decisions and find that fake audio detectors prioritize the artifacts of concatenated parts (also known as transition regions). WeDefense integrates Grad-CAM to facilitate such interpretability, helping researchers assess and improve the reliability of their models.

\subsection{Devployment} 
WeDefense supports exporting models to Just In Time (JIT). Meanwhile, it offers Command Line Interface (CLI) that is accessible through a straightforward ``pip
install'' process. Thus models from WeDefense can be easily adopted in the deployment environment. WeDefense also provides interactive demos for fake audio detection\footnote{\url{https://huggingface.co/spaces/wedefense/fake_audio_detection_demo}} and localization in huggingface\footnote{\url{https://huggingface.co/spaces/wedefense/fake_audio_localization_demo}}, the demo page is illustrated in Figure. \ref{fig:demo_localization}.

\section{Recipes and Results on Detection}

\subsection{Configuration}
To save computational resources, we follow several common configurations unless otherwise specified:
(1). All audio segments are truncated or padded to 3 seconds during training.
(2). Except for experiments reported in Table~\ref{tab:detection_augmentation}, no data augmentation is applied.
(3). ResNet18 and LCNN-LSTM are trained for 100 epochs, with the final model obtained by averaging the last 10. SSL-based models are trained for 5 epochs, averaging the last 2.

More configuration details can be found in the YAML files available in the WeDefense repository\footnote{The repository will be made available after the review period.}. It is important to note that, although we adopt simplified settings for efficiency in this paper, all previously mentioned features and options in Section~\ref{sec:wedefense} are fully supported by the WeDefense toolkit.

\subsection{Model Comparison}\label{sec:detection_models_comparison}
In this section, we select two model categories from WeDefense those commonly used in fake audio detection:
\begin{itemize}
\item \textbf{Non-SSL CNN-based models}: ResNet18 and LCNN-LSTM. ResNet18 was the most widely adopted model among participants in the close condition of the latest ASVspoof5 challenge~\cite{rohdin24_asvspoof, duroselle24_asvspoof, chan24_asvspoof}. LCNN-LSTM is included due to its good performance in anti-spoofing tasks.

\item \textbf{SSL-based Models}: 
Recent studies suggest that a complex backend is no longer necessary~\cite{zhang2024sls}. Therefore, we evaluated the widely used SSL-AASIST~\cite{tak22_odysseyw2v2} and several lightweight backends: SLS~\cite{zhang2024sls}, MHFA~\cite{Peng2023AnVerification},  Res1D~\cite{liu24grad_cam}, gMLP~\cite{Liu2021gmlp}.
For PartialSpoof, we adopt XLSR-53 as the front-end SSL model, as it outperforms other SSL models under an ablation study using SSL-MHFA\footnote{Detailed results can be found in the repo.}.
For ASVspoof5, we adopt wav2vec2\_large\_960 as it is the most widely adopted model among participants in the open condition of the latest ASVspoof5 challenge~\cite{kulkarni24_asvspoof, xu24_asvspoof, zhu24_asvspoof}. For SSL-based models, both the SSL-based frontends and backends are updated in our experiments.
\end{itemize}

Results on PartialSpoof and ASVspoof5 are shown in Table~\ref{tab:detection_partialspoof} and \ref{tab:detection_ASVspoof5} respectively.
The best performance reported on PartialSpoof is included for reference. 
 Unsurprisingly, SSL-based models outperform non-SSL CNN-based models such as ResNet18 and LCNN-LSTM, benefiting from pretraining on large-scale datasets and the powerful representation capabilities of transformer layers. Notably, we observe that with SSL-based front-ends, even light-weight backends or simple backend with fusion structure over transformer layers can still achieve strong performance. 

\begin{table}[!t]
\caption{EER (\%) of Fake detection on the partialspoof. (Prefix of IDs refers to the training data, and suffixes are following the name in the repo that starts with v(ersion).)}\label{tab:detection_partialspoof}
\vspace{-2mm}
\centering
\begin{tabular}{ccccc}
\toprule
\textbf{ID}  & \textbf{Frontend} & \textbf{Backend}   & \textbf{Dev.}  & \textbf{Eval.} \\ \midrule
PS\_v03 & \multirow{2}{*}{Fbank}             & ResNet18  & 4.196 & 6.921 \\
PS\_v01 &              & LCNN-LSTM & 3.022 & 5.898 \\ \midrule
PS\_v14 &  \multirow{5}{*}{xlsr\_53}          & SLS       & 0.834 & 1.794 \\
PS\_v11 &           & gMLP      & 0.668 & 1.384 \\
PS\_v12 &          & Res1D     & 0.504 & 1.414 \\
PS\_v13 &           & AASIST      & 0.551 & 1.044 \\
PS\_v15 &           & MHFA      & 0.237 & 0.802 \\
\midrule
 \cite{liu2025nes2net} & wav2vec2\_large  & Nes2Net & 0.33\phantom{0} & 0.64\phantom{0}  \\
\bottomrule
\end{tabular}
\end{table}

\begin{table}[!t]
\caption{EER (\%) of Fake detection on the ASVspoof5.}\label{tab:detection_ASVspoof5}
\vspace{-2mm}
\centering
\begin{tabular}{ccccc}
\toprule
\textbf{ID}  & \textbf{Frontend} & \textbf{Backend}   & \textbf{Dev.}  & \textbf{Eval.} \\
\midrule
A5\_v01 & \multirow{2}{*}{Fbank}             & LCNN-LSTM     & 21.810 & 50.432 \\
A5\_v03 &              & ResNet18      & 18.619 & 32.572 \\ \midrule
A5\_v13 & \multirow{4}{*}{wav2vec2\_large\_960} & AASIST      & \phantom{0}6.680  & 17.113 \\
A5\_v14 &  & SLS         & \phantom{0}1.740  & 12.151 \\
A5\_v12 &  & Res1D       & \phantom{0}5.205  & 11.177 \\
A5\_v15 &  & MHFA        & \phantom{0}1.235  & \phantom{0}8.083  \\
\bottomrule
\end{tabular}
\end{table}

As shown in Table~\ref{tab:detection_partialspoof}, on the small dataset as PartialSpoof, there are not many differences between different backends no matter whether they are complicated backends or simply attention fusion of transformer outputs. After utilizing SSL, the system can achieve relatively good performance under suitable parameter settings. And v15 implemented in WeDefense achieves performance comparable to the best results reported in the literature. 
%
For ASVspoof5 as shown in Table~\ref{tab:detection_ASVspoof5}, the v01 (LCNN-LSTM) failed on the evaluation set, due to the presence of adversarial attacks specifically targeting LCNN-LSTM~\cite{todisco24malacopula}. 


\subsection{Augmentation Comparison}
Augmentation is proven to be an efficient way in the latest ASVspoof5 challenge~\cite{duroselle24_asvspoof}. In this section, we compared those mostly applied augmentation methods on the ASVspoof5. Results are shown in Table~\ref{tab:detection_augmentation}. And the best performed model is also included, which incorporates a more complex augmentation.
WeDefense achives slight better performance using only MUSAN and RIRs for fusion.
We can see that although some of the augmentations improve performance, others, like speed perturb, may degrade it with increased EER. This suggests we should be careful when applying augmentation to training models.

\begin{table}[!h]
\caption{Comparison of Different Augmentation Methods and the State-of-the-Art Systems on ASVspoof5}\label{tab:detection_augmentation}
\vspace{-2mm}
\centering
\begin{tabular}{cccc}
\toprule
\textbf{ID}  & \textbf{Augmentation} & \textbf{Dev.} & \textbf{Eval.} \\
\midrule
Aug 0 & w/o aug               & 1.235 & \phantom{0}8.083 \\
\midrule
Aug 1 & speed perturb         & 4.883 & 14.868 \\
Aug 2 & codec                 & 0.696 & \phantom{0}7.009 \\
Aug 3 & rawboost              & 1.711 & \phantom{0}5.091 \\
Aug 4 & musan + rirs          & 0.648 & \phantom{0}4.821 \\
\midrule

\multicolumn{2}{c}{\cite{falez24_asvspoof} wav2vec2\_large\_960 + RawGAT-ST + aug.} & 0.76\phantom{0} & -- \\

\multicolumn{2}{l}{\cite{zhu24_asvspoof} SLIM - Best single system in ASVspoof5} & 3.00\phantom{0} &\phantom{0}5.56\phantom{0}  \\
\bottomrule
\end{tabular}
\end{table}

\subsection{Calibration} 
This subsection compares the impact of applying calibration. Specifically, LLRs (log-likelihood ratio) are calibrated with logistic regression~\cite{ferrer2024calibration}, which is trained on the ASVspoof5 development set.
Results of w/o and w/ applying LR-based calibration are shown in Table \ref{tab:ASVspoof5_calibration}.

\begin{table}[!t]
\caption{Impact of calibration on the ASVspoof5 dataset.}\label{tab:ASVspoof5_calibration}
\vspace{-2mm}
\centering
\scalebox{0.97}{
\begin{tabular}{cccccc}
\toprule
                          &                & \multicolumn{2}{c}{\textbf{Dev.}} & \multicolumn{2}{c}{\textbf{Eval.}} \\ \cmidrule(lr){3-4}\cmidrule(lr){5-6}
\textbf{Model}&\textbf{Calibration}    & \textbf{$C_\text{llr}$}       & \textbf{actDCF}      & \textbf{$C_\text{llr}$}        & \textbf{actDCF}      \\
\midrule
\multirow{2}{*}{ResNet18 (v03)}         & w/o & 4.143  & 0.741  & 5.919 & 0.887 \\
                                  & LR & 0.538  & 0.283  & 0.914 & 0.723 \\
\midrule
\multirow{2}{*}{SSL-MHFA (v15)}  & w/o & 0.403 & 0.121 & 1.279 & 0.417  \\
                           & LR & 0.076 & 0.047 & 0.564 & 0.224  \\
\bottomrule
\end{tabular}}
\end{table}

As can be seen in Table \ref{tab:ASVspoof5_calibration}, the calibration sensitive metrics $C_\text{llr}$ and $\text{actDCF}$ improves substantially by LR-based calibration. It should be noted that the \emph{w/o calibration} scores have been obtained by converting the posteriors from the neural networks' softmax layer into LLRs, taking the ratio of bonafide/spoofed utterances in the training data into account. This should lead to calibrated scores in principle. However, the calibration may often be destroyed in practice due to, e.g., overfitting or mismatch between the properties of the training and the test data.

\subsection{Fusion}
Fusion is a common strategy used to ensemble advantages from different systems. It is especially popular in the challenges. In this subsection, we compare two score-level fusion methods for the cross-database scenario: score averaging and LR-based fusion.
For each row, the fusion model in each evaluation set (column) is not the same but its fusion parameters (e.g., normalization or LR weights) are estimated on the development set corresponding to each evaluation set.
For the score averaging, the development set is used to normalize the scores of the individual systems into the $[0-1]$ range. For LR-based fusion, parameters of the LR model are trained on the development set. The LR model is the same as the one used for calibration in the previous subsection. 

\begin{table}[!t]
\caption{EER (\%) of Fusion on the Cross-Database Scenario. (``PS\_'' denotes models trained on PartialSpoof, and ``A5\_'' denotes models trained on ASVspoof5.)}\label{tab:fusion}
\centering
\begin{tabular}{cccccc}
\toprule
                         &         & \multicolumn{2}{c}{\textbf{PartialSpoof (PS)}} & \multicolumn{2}{c}{\textbf{ASVspoof5 (A5)}} \\ \cmidrule(lr){3-4}\cmidrule(lr){5-6}
                         &         & \multicolumn{1}{c}{\textbf{Dev.}} & \textbf{Eval.} & \textbf{Dev.}          & \textbf{Eval.}         \\
\midrule
\multirow{2}{*}{-} & PS\_v15 &   \phantom{0}0.237  & \phantom{0}0.802  & 16.755 & 34.513 \\
                         & A5\_v15            & 18.613 & 20.216 & \phantom{0}1.235  & \phantom{0}8.083   \\
\midrule
\multirow{3}{*}{Average} & PS\_(v13+v15)      &  \phantom{0}0.274 & \phantom{0}0.843    & 10.982           &    27.151           \\
                         & A5\_(v14+v15)      &  15.700 & 16.046   &  \phantom{0}0.986          &   \phantom{0}8.011       \\
                         & PS\_v15 + A5\_v15  &  \phantom{0}0.243  & \phantom{0}0.886    &  \phantom{0}4.909        &      10.251      \\
\midrule
\multirow{3}{*}{LR}      & PS\_(v13+v15) &  \phantom{0}0.239  &  \phantom{0}0.762    &    \phantom{0}7.337  & 21.721              \\
                         & A5\_(v14+v15) &  15.981 & 16.545                          &  \phantom{0}1.005 & \phantom{0}8.371            \\
                         & PS\_v15 + A5\_v15  &  \phantom{0}0.241 & \phantom{0}0.870 &    \phantom{0}1.273 & \phantom{0}8.097              \\

\bottomrule
\end{tabular}
\end{table}
Results are shown in Table \ref{tab:fusion}. We can see that cross-database evaluation reveals that models trained on one dataset perform poorly on a different unseen dataset.
However, this performance degradation can be mitigated to some extent using simple fusion strategies, without requiring retraining on combined data. 
Fusion strategies can be applied in two ways, depending on available databases:
\begin{itemize}
    \item Single-database model fusion (Rows 3–4 and 7–8): Fusing different models trained on the same database improves robustness.
    \item Cross-database model fusion (Rows 5 and 9): When models trained on both databases are available, fusing them reaches similar performance as using only the relevant single database model, i.e., the irrelevant model has effectively been discarded by the fusion.  
\end{itemize}

\subsection{Analyses} 
\subsubsection{Embedding Visualization - UMAP}

\begin{figure}[!tb]
\centerline{\includegraphics[width=0.7\linewidth]{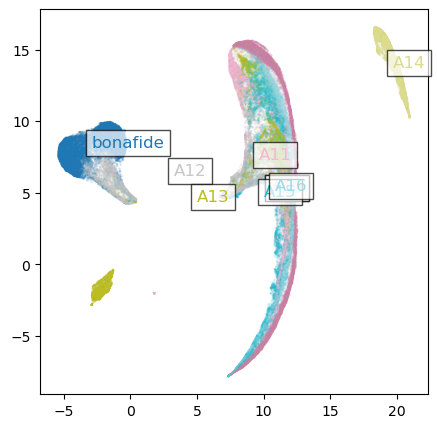}}
\vspace{-2mm}
\caption{Example of Embedding Visualization by UMAP  on the dev. set of ASVspoof5 (annotated according to the released metadata).}\label{fig:umap}
\end{figure}

Figure~\ref{fig:umap} shows the embedding extracted from the development set of ASVspoof5 by using A5\_v15, we can see that samples from A12 (In-house unit-selection) in gray are mixed with real samples in blue. These samples are typically generated by concatenating selected segments from bona fide utterances using a unit-selection approach and were not seen during training. This observation aligns with the findings reported for ResNet18 in~\cite{rohdin24_asvspoof}.

\subsubsection{XAI - Grad-CAM}

\begin{figure}[!tb]
\centerline{\includegraphics[width=0.8\linewidth]{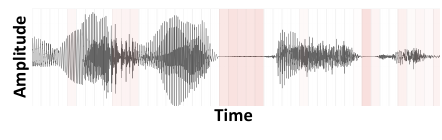}}
\caption{Frame-level Grad-CAM visualization of parts of a segment from CON\_E\_0033629.wav in the PartialSpoof evaluation set. Regions with darker red indicate higher model contribution.}
\label{fig_xai}
\vspace{-2mm}
\end{figure}

Previous work on XAI investigated how neural countermeasures detect partially spoofed speech using Grad-CAM~\cite{liu24grad_cam}. 
We re-implement and integrate this capability into WeDefense. 
Following their work~\cite{liu24grad_cam}, we applied Grad-CAM to SSL-Res1D model. An example of Frame-level Grad-CAM visualization is shown in Figure~\ref{fig_xai}. And the Relative Contribution Quantification (RCQ) values on the PartialSpoof are shown in Table~\ref{tab_xai}.
Our visualization and RCQ results align with their original findings~\cite{liu24grad_cam}, showing that the concatenated parts consistently exhibit the highest RCQ, which indicates a higher contribution to the model detection decision. 

\begin{table}[]
\centering
\label{tab_xai}
\caption{RCQs (\%) of the SSL-Res1D model (PS\_v12) when predicting scores for partially spoofed samples on the PartialSpoof Dev. and Eval. sets, broken down by five different segment types.}
\setlength{\tabcolsep}{1.33mm}{
\begin{tabular}{cccccc}
\toprule
RCQ (\%) & \begin{tabular}[c]{@{}c@{}}Bona fide\\ Speech\end{tabular} & \begin{tabular}[c]{@{}c@{}}Spoofed\\ Speech\end{tabular} & \begin{tabular}[c]{@{}c@{}}Concatenated\\ Parts\end{tabular} & \begin{tabular}[c]{@{}c@{}}Bona fide\\ Non-speech\end{tabular} & \begin{tabular}[c]{@{}c@{}}Spoofed\\ Non-speech\end{tabular} \\ 

\midrule
Dev. Set & -38.38 & -13.78 & \textbf{57.47} & 30.91 & 46.65 \\
Eval. Set & -31.40 & -12.07 & \textbf{59.70} & 40.53 & 35.58 \\ \bottomrule
\end{tabular}
}
\end{table}

\section{Recipes and Results on Localization}
The current version of WeDefense also supports frame-level fake audio localization. This is primarily achieved through uniform segmentation methods, which are straightforward and enable frame-wise predictions. 
Specifically, this can be done by adapting the detection models introduced in Section~\ref{sec:detection_models_comparison} and removing the pooling layers of backends. In addition, we integrate another uniform segmentation method, SSL-BAM~\cite{zhong2024BAM} model into WeDefense, which leverages explicit boundary information for localization.

\subsection{Configuration}
Following our finding in Section~\ref{sec:detection_models_comparison} that SSL-based front-ends combined with lightweight back-ends can deliver strong performance, we implement localization variants of SSL-MHFA and SSL-SLS by removing their pooling layers to support uniform frame-level segmentation. Consistent with the detection setup, we use xlsr\_53 as the front-end for all SSL-based localization models discussed in this section, including SSL-MHFA, SSL-SLS, and SSL-BAM. Detailed configurations can be found in the repository.

\subsection{Results Comparison}
Results on fake audio localization are shown in Table~\ref{tab:localization}.
We also included results from a pre-trained model called CFPRF~\cite{wu2024CFPRF} for reference. 
It demonstrates the best performance on the evaluation set of PartialSpoof and we are working on integrating it into WeDefense in future updates.

Due to the presence of numerous short fake segments in PartialSpoof, the localization task remains highly challenging. The EERs on the evaluation set are still relatively high, highlighting the need for further research in this direction.

\begin{table}[!h]
\caption{EER (\%) for fake audio localization on the PartialSpoof dataset.}
\label{tab:localization}
\centering
\begin{tabular}{ccc}
\toprule
\textbf{Models} & \textbf{Dev.} & \textbf{Eval.} \\
\midrule
xlsr\_53 + MHFA & 1.35 & 12.13 \\
xlsr\_53 + SLS\phantom{Xa}  & 1.66 & 21.57 \\
xlsr\_53 + BAM\phantom{X} & 4.84 & 12.53 \\
\midrule
\cite{wu2024CFPRF} CFPRF (xls-r 300m\cite{babu22_interspeech} + boundary matching) & 1.90  & \phantom{0}7.72 \\
\bottomrule
\end{tabular}
\end{table}

\section{Conclusion and Future Work}
In this paper, we describe WeDefense, the first open-source toolkit for fake audio detection and localization. 
WeDefense is a comprehensive toolkit that covers every step potentially involved in defending against fake audio. It includes both detection and localization tasks at the time of writing this paper.
WeDefense is regularly updated, and the current development addresses the following areas.
\begin{enumerate}
\item For the current toolkit version, we standardized key parameters across different models to focus on the evaluation of different models and architectures, instead of replicating strictly the configurations reported in original research articles.  As a result, performance estimates reported in this article may differ from those reported in the original work. We are currently working to optimize the training strategies in particular and plan to release updates in the near future.
\item  Currently, the number of supported models, especially for localization, is modest. WeDefense is nonetheless an active project which continues to attract broader adoption and contributors. The integration of emerging, more advanced detection and localization models is a priority.
\end{enumerate}

We hope that WeDefense will lower the cost of entry to the fake audio detection and localization research and that it will help to attract more researchers and contributors to the field. In time, we also hope a growing community can begin to close the gap between progress in fake audio detection and generative speech technology.

\balance
\bibliographystyle{IEEEtran}
\bibliography{main}

\end{document}